\documentclass[twocolumn]{revtex4-1}
\usepackage{graphics}
\usepackage{amsmath}
\usepackage{amssymb}
\usepackage{amsfonts}

\begin{document}

\title{Resonant Generation of an Electron–Positron Pair\\
by Two Photons to Excited Landau Levels}

\author{M. M. Diachenko} 
\email{dyachenko.mikhail@mail.ru}

\author{O. P. Novak}
\email{novak-o-p@ukr.net}

\author{R. I. Kholodov}
\email{kholodovroman@yahoo.com}

\affiliation{The Institute of Applied Physics of National Academy of Sciences of Ukraine, 
58, Petropavlivska Street, 40000, Sumy, Ukraine}


\begin{abstract} 
We consider the resonant generation of an electron–positron pair by two polarized photons to
arbitrarily low Landau levels. The resonance occurs when the energy of one photon exceeds the one-photon
generation threshold, and the energy of the other photon is multiple to the spacing between the levels. The
cross section of the process is determined taking into account the spins of particles. The order of magnitude
of the cross section is the highest when the magnetic moments of the particles are oriented along the magnetic
field.

\vspace{1ex}
\noindent
\textit{J. Exp. Theor. Phys.~\textbf{121}(5) 813 (2015);}  \quad \textbf{doi:} 10.1134/S1063776115110126 
\end{abstract}

\maketitle

\section{Introduction}

Processes accompanying ion collisions continue toattract considerable attention due to the advances in
accelerator technology. 
In particular, a new Facility for Antiproton and Ion Research (FAIR) is being constructed based on the GSI Helmholtz Center for Heavy Ion Research at Darmstadt, Germany ~\cite{FAIR}.

Experiments with heavy ions open wide prospects for the verification of quantum electrodynamics in strong magnetic fields. 
The generation of an electron-positron pair upon collisions of ions is one of the most interesting processes. This process was considered for the first time by Landau and Lifshitz~\cite{Landau34} and was later studied in detail in the high energy range as well as for low-energy collisions (see, for example, \cite{Hencken95, Baur02, Baur07, Arbuzov11,    Reinhardt81, Greiner85, Nehler94, McConnell12} and references therein).

In the latter case, a quasi-molecule can be formed.
The magnetic field produced by moving ions can reach and even exceed the critical quantum-electrodynamics value of $H_c \approx 4.41\cdot 10^{13}$~G even for ion energies on the order of the Coulomb barrier. 
Nevertheless, the effect of the magnetic field is usually disregarded in accordance with the conclusions drawn in~\cite{Soff81, Rumrich87, Soff88}.

However, as shown in~\cite{Fomin98}, the magnetic field can produce a substantial effect due to the interaction with the generated pair. 
The mechanism of such an interaction is analogous to magnetic field line freezing into a plasma (this effect is well known in the plasma physics and astrophysics). 
As a result, the lifetime of the magnetic field increases significantly and can considerably exceed the nuclear time of flight. 
An observable indication of this effect is the presence of resonances in the spectrum of generated pairs, which are typical of processes occurring in a magnetic field. 
It is interesting to note that anomalous peaks were observed in GSI experiments on collisions of heavy ions~\cite{Backe78, Cowan86, Koenig87}, although the reproduction of the results unfortunately remains problematic.


We believe that the main features of this effect can be investigated using the following model. 
The formation of a pair can be described as a photoproduction by two photons using the well-known equivalent-photon approximation. 
The corrections corresponding to the interaction with the magnetic field can be taken into account in the context of the Furry picture (i.e., using the electron wavefunctions in a magnetic field). 
It should be noted that this technique is generally analogous to the Bethe–Maximon method used for describing collisions of ions with relativistic energies~\cite{Bethe54, Davies54}. 
Therefore, in the simplest case, the problem can be reduced to two-photon generation of a pair in a magnetic field.


This problem is also of independent importance in astrophysics. In particular, the effectiveness of generation of an electron–positron plasma in the magnetospheres of pulsars by competing one- and two-photon processes was considered in~\cite{Burns84,  Zhang01, Harding02}.


The process of generation of an electron–positron pair by two photons in a strong magnetic field was studied in~\cite{Ng77} in the case of a head-on collision of photons along the magnetic field line. 
This process was analyzed in~\cite{Kozlenkov86} in the nonresonant case, when the energy of each of the photons is insufficient for generating a pair in a one-photon process. 
The mean free path of a high-energy photon propagating through a photon gas along magnetic field lines was calculated in~\cite{Dunaev12}.


In this study, we consider the effect of the resonant production of an electron–positron pair by two polarized photons in a magnetic field on the excited Landau levels taking into account the spins of the particles. 
We calculate the resonant cross section for arbitrary polarizations of particles in the case when the electron and the positron occupy arbitrary low Landau levels.

We use the relativistic system of units ($\hbar = c = 1$).



\section{Probability amplitude and kinematics of the process}

The expression for the amplitude of two-photon production of an electron–positron pair in a magnetic field has the form
%
\begin{gather}
  \nonumber
  S_{fi} = -ie^2 \int d^4x d^4x' 
  \left[
  \bar\Psi_e (A_1\gamma) G(x-x') (A'_2\gamma) \Psi_p' +
  \right.
  \\ \label{Sfi}
  \left.
  + \: \bar\Psi_e (A_2\gamma) G(x-x') (A'_1\gamma) \Psi_p'
  \right],
\end{gather}
where $A_{1,2}$ are the potentials of plane waves~\cite{LandauIV} and $\Psi_{e,p}$ are the solutions to the Dirac equation in a magnetic field for the electron and positron (primed values are functions of $x'$). 
We direct the $z$ axis along the magnetic field and choose the vector potential in the form $\vec A = (0,xH,0)$. 
Then the electron wavefunctions describe the states with certain values of energy and momentum~\cite{Akhiezer}. 
The energy eigenvalues are given by
%
\begin{equation}
  \label{El}
  E_l = \sqrt{\tilde{m}^2 + p_z^2}, \qquad
  \tilde{m} = m\sqrt{1 + 2lh},
\end{equation}
where $m$ is the electron mass, $l$ is the number of the Landau level, and $h$ is the magnetic field strength in units of the critical field:
%
\begin{equation}
  h = H/H_c.
\end{equation}

The electron propagator in the given basis has the
form~\cite{Fomin99}
%
\begin{equation}
\label{G}
  G(x-x') = \frac{-m\sqrt{h}}{(2\pi)^3} \int d^3g e^{-i\Phi} \sum\limits_{n} \frac{G_H(x,x')}{g_0^2 - E_n^2},
\end{equation}
\begin{equation}  
  \label{GH}
  \begin{array}{l}
    G_H(x,x') = U_n(\rho) U_n(\rho') \: (\gamma P + m) \: \tau + 
    \\
    + im \sqrt{2nh} \:  U_{n-1}(\rho)U_{n}(\rho') \: \gamma^1\tau - 
    \\ 
    - im \sqrt{2nh}  \: U_n(\rho) U_{n-1}(\rho')  \: \tau \gamma^1 +
    \\
    + U_{n-1}(\rho) U_{n-1}(\rho')  \: (\gamma P + m) \: \tau^*,
  \end{array}
\end{equation}
where  $d^3g = dg_0 dg_y dg_z$,  $\gamma$ are the Dirac gamma matrices,
\begin{gather}
  \Phi = g_0(t-t') - g_y (y-y') - g_z (z-z'),\\
  \tau = \frac{1}{2} (1 + i \gamma_2\gamma_1), \\
  E_n = \sqrt{m^2 + g_z^2 + 2nh m^2}, \\
  P = (E_n, 0, 0, g_z).
\end{gather}
The argument of Hermite functions $U_n(\rho)$ in Eq.~(\ref{GH}) has the form
%
$$
  \rho = m\sqrt{h} x + g_y/m\sqrt{h},
$$
It should be noted that propagator (\ref{G}) was also obtained independently in \cite{Kuznetsov11}

\begin{figure}
  \resizebox{\columnwidth}{!}{\includegraphics{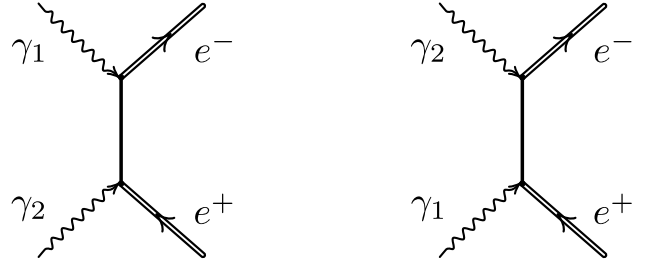}}
  \caption{Feynman diagrams of production of an electron–
positron pair by two photons.}
  \label{diagrams}
\end{figure}

Figure~\ref{diagrams} shows the Feynman diagrams corresponding to amplitude (\ref{Sfi}). 
The amplitude was obtained in explicit form in~\cite{Diachenko14}.

The kinematics of the process is determined by the law of conservation of energy and the $z$ component of the momentum,
\begin{equation}
  \label{laws}
  \begin{array}{l}
    \omega_1 + \omega_2 = E^- + E^+, \\
    k_{1z} + k_{2z} = p_z^- + p_z^+.
  \end{array}
\end{equation}
%
(here and below, superscripts ``$+$'' and ``$-$'' correspond to the positron and electron, respectively).
It can easily be verified~\cite{Diachenko14} that the threshold frequencies and momenta of photons satisfy the condition
\begin{equation}
\label{th}
  \left( \omega_1^{th} + \omega_2^{th} \right)^2 -
  \left( k_{1z}^{th} + k_{2z}^{th} \right)^2 =
  \left( \tilde{m}^- + \tilde{m}^+ \right)^2 .
\end{equation}

It can be seen that this conditions cannot be satisfied if both photons move parallel to the field in the same direction. 
In this case, the left-hand side of Eq.~(\ref{th}) is zero, while the right-hand side is always greater than $(2m)^2$.

It should be noted that the Lorentz transformations along the magnetic field do not change the field itself.
Therefore, without loss of generality, we can eliminate the longitudinal momentum of the photons by appropriately choosing the reference system so that 
\begin{equation}
\label{K0}
  k_{1z} + k_{2z} = 0.
\end{equation}
%
In addition, we consider the process in the ultra-quantum-mechanical, or low Landau levels (LLL) approximation, in which the following conditions hold:
\begin{equation}
\label{LLL}
  hl^\pm \ll 1, \qquad
  l^\pm \sim 1.
\end{equation}
%

We introduce detuning $\delta \omega$ from the threshold,
\begin{equation}
  \label{dw}
  \delta\omega = \omega - (\tilde{m}^- + \tilde{m}^+),
\end{equation}
%
where we have introduced the notation $\omega = \omega_1 + \omega_2$.
We assume that the detuning is on the order of the spacing between the Landau levels $\delta \omega \sim mh$. 
Taking into account conditions (\ref{LLL}), we can then write the momentum of the generated particles as
\begin{equation}
\label{pzlll}
  |p_z^\pm| \approx \sqrt{m\delta\omega}.
\end{equation}

\section{Conditions for resonant process}
When quantities $g_0$ and $g_z$ in propagator (\ref{G}) satisfy the relativistic relation between the energy and momentum of a particle in a magnetic field, resonance takes place. 
The resonance condition is the equality to zero of the denominator of the Green’s function:
%
\begin{equation}
\label{res}
  g_0 = \pm E_n. 
\end{equation}
Quantities $g_0$ and $g_z$ are defined in accordance with the laws of conservation at the vertices of the diagram.

Equations (\ref{res})  define two resonance conditions.
We will use conditions (\ref{LLL}) and expression (\ref{pzlll}) for the momentum of finite particles. 
In this case, resonance frequencies of photons in the lowest approximation are given by
%
\begin{equation}
\label{resw1}
  \left\{
  \begin{array}{l}
    \omega_1^{res} \approx mh (l^- - n),\\
    \omega_2^{res} \approx 2m + mh (l^+ + n) + \delta\omega,
  \end{array}
  \right.
\end{equation}
\begin{equation}
\label{resw2}
  \left\{
  \begin{array}{l}
    \omega_1^{res} \approx 2m + mh (l^- + n) + \delta\omega, \\
    \omega_2^{res} \approx mh (l^+ - n),
  \end{array}
  \right.
\end{equation}
for the ``$+$'' and ``$-$'' signs in Eq.~(\ref{res}), respectively. 
It should be noted that the intermediate particle is the electron in the former case and the positron in the latter case.

It is well known that a second-order process in the fine-structure constant at resonance can be depicted as a sequence of first-order processes. 
In our case, this is the generation of a pair by a single photon and absorption of the photon in the magnetic field. 
It follows from Eqs. (\ref{resw1}) and (\ref{resw2}) that one of the photons (the hard one) forms a pair and should have an energy exceeding the one-photon generation threshold. 
The second (soft) photon is absorbed by the electron, and its frequency should be equal to the energy of transition between Landau levels (i.e., multiple to cyclotron frequency $\omega_H = mh$).

The resonance conditions in the Feynman exchange diagram can be obtained from expressions (\ref{resw1}) and (\ref{resw2}) by interchanging photon indices $1\rightleftarrows 2$. 
In the general case, number $n$ also changes $n\rightarrow n'$. 
We can obviously formulate the conditions in which resonance is possible for two diagrams simultaneously. 
Indeed, equating the frequencies of soft photons in the direct and exchange diagrams, we obtain
%
\begin{equation}
\label{interf}
  l^- - n = l^+ - n'.
\end{equation} 

Let us formulate the condition for interference of resonances in a more convenient form. 
We can write the energy of the soft photon in resonance as
\begin{equation}
  \label{N}
  \omega^{res} = mh \cdot N, \qquad N = 1,2,...
\end{equation}
%
Then, simultaneous resonance of the amplitudes is possible for values of $n$ and $n'$ given by
\begin{equation}
  \label{intN}
  \left\{
  \begin{array}{l}
    n  = l^- - N,\\
    n' = l^+ - N.
  \end{array}
  \right.
\end{equation}
Obviously, interference of resonances is impossible if $l^- < N$ or $l^+ < N$. 
Indeed, the soft photon in resonance is absorbed by different particles (electron and positron) in the direct and exchange diagrams. 
If the photon energy is higher than the energy of one of the particles, the resonance condition for the corresponding amplitude is violated.

Henceforth, we will mark the quantities pertaining to the soft photon with the subscript ``1''. 
In addition, we confine our analysis to the case when resonance occurs only in one diagram. 
For definiteness, we set $N > l^+$.

In the first nonvanishing approximation in $h$, the resonance frequencies are independent of the angles of incidence of photons. 
Calculation of the next correction in $h$ gives
%
\begin{equation}
  \label{reswu}
  \begin{array}{l} 
    \delta\omega_1^{res}  = \pm \cos\theta_1 \: \omega_1^{res}  \sqrt{\delta\omega/m},\\
    \delta\omega_2^{res}  = - \delta\omega_1^{res} .
  \end{array}
\end{equation}
%
where $\theta_1$ is the polar angle of the first photon, and the sign in the first expression coincides with the sign $p_z^-$.

It should be noted that the hard resonant photon must be directed almost perpendicularly to the magnetic field in view of the choice of reference system (\ref{K0}). 
Using relation (\ref{resw1}), we obtain
%
\begin{equation}
  \cos\theta_2 \approx - \frac{hN}{2} \cos\theta_1.
\end{equation}

\begin{figure}
  \resizebox{\columnwidth}{!}{\includegraphics{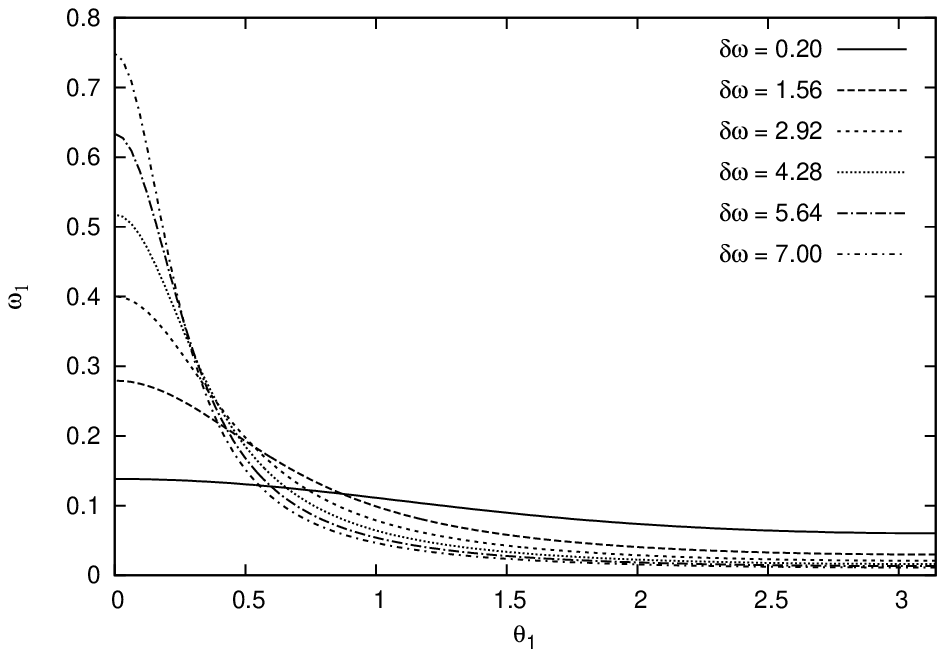}} 
  \\ 
  \resizebox{\columnwidth}{!}{\includegraphics{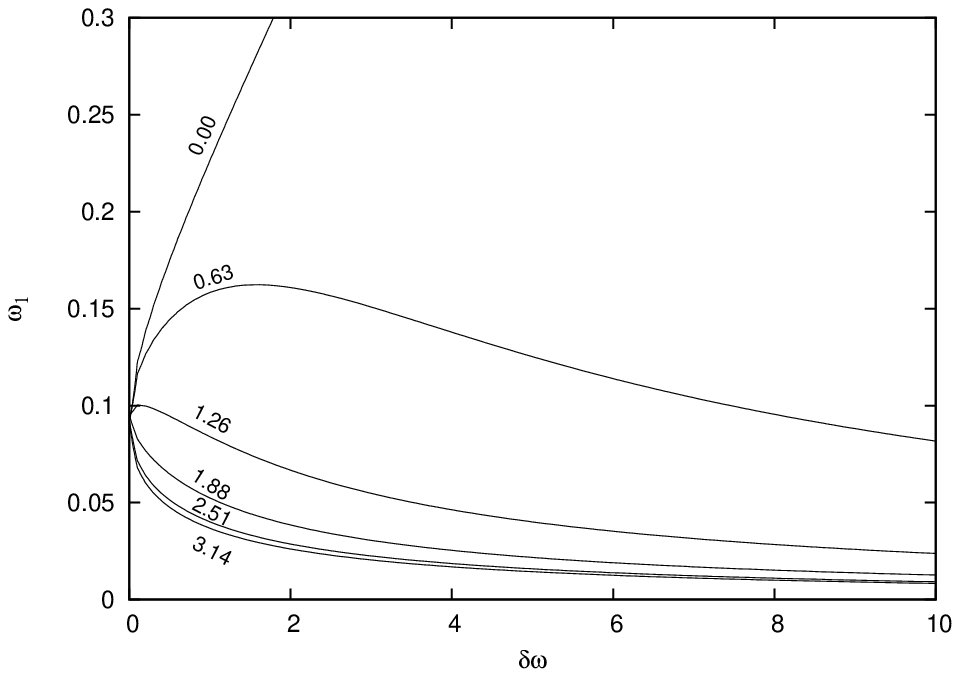}}
  \caption{Exact dependences of resonance frequency $\omega_1$ of the soft photon on (a) its polar angle $\theta_1$ and (b) detuning $\delta\omega$ for levels $l^- = 1$ and $l^+=0$, and magnetic field $h = 0.1$.}
  \label{t_w1}
\end{figure}

Figure~\ref{t_w1} shows the dependences of resonance frequencies of photons, which are exact in $h$.

%
%
%

\section{Resonant cross section of the process}
Let us calculate the resonant cross section of the process using expressions (\ref{resw1}) for the resonance frequencies and assuming that conditions (\ref{LLL}) is satisfied.

It is well known that the differential cross section of the process has the form
%
\begin{equation}
\label{dWgen}
  d\sigma = \frac{\left| S_{fi} \right|^2}{1 - \cos\chi} V \: dN^- dN^+, 
\end{equation}
%
where $dN^\pm$ are the intervals of final states of the electron and positron,
\begin{equation}
  dN^\pm = \frac{S d^2p^\pm}{(2\pi)^2},
\end{equation}
$S$ and $V$ are the area and volume of the normalized wavefunctions for the electron and photon, $d^2p^\pm = dp^\pm_y dp^\pm_z$, and  $\chi$ is the angle between the directions of the photons.

Substituting the explicit form of the wavefunctions into expression (\ref{Sfi}) and retaining only the first terms in the expression in small parameter $h$, we obtain 
%
\begin{align}
\label{dsigma}
  d\sigma = \frac{2\pi\alpha^2}{hN}
             \frac{q_1^Nq_2^{l^++n}}{\left| g_0^2 - \varepsilon_n^2 \right|^2}
             \frac{S e^{-q_2}}{V(1-\cos\chi)}
             \frac{l^-!/l^+!}{(N!n!)^2} 
             \times \\ \times \nonumber
             |A_{\mu^-\mu^+}|^2 \:\delta^3\: d^2p^- d^2p^+
\end{align}
where $\alpha$ is the fine-structure constant, and indices $\mu^\pm$ denote the signs of the spin projections of the electron and position, respectively. 
The spin-dependent factors have the form
%
\begin{equation}
  A_{-+} = e^{i\phi_1}\sqrt{2}\frac{e_{2z}T_1^-}{\sin\theta_1},
\end{equation}
\begin{align}
    A_{++} = \sqrt{\frac{h}{l^-}}\left[
             Ne_{1z}e_{2z} - Ne_{2z}T_1^-e^{i\phi_1}\frac{\cos\theta_1}{\sin\theta_1} +
             \right. \\ \nonumber \left. 
             + n\frac{T_1^-T_2^-}{\sin\theta_1\sin\theta_2}
             \right] ,
\end{align}
\begin{equation}
    A_{--} = e^{i(\phi_2-\phi_1)}\sqrt{hl^+}\frac{T_1^-T_2^+}{\sin\theta_1\sin\theta_2} , 
\end{equation}
\begin{align}
    A_{+-} = e^{-i\phi_2}\sqrt{\frac{h^2l^+}{2l^-}}  \left[
             ne^{i(\phi_2+\phi_1)}e_{2z}T_1^-\frac{\mbox{ctg}^2\theta_2}{\sin\theta_1} + 
             \right.
             \\ \nonumber
             \left.
             + Ne^{i\phi_1} T_1^-T_2^+ \frac{\mbox{ctg}\theta_1}{\sin\theta_2}
             - N \frac{e_{1z}T_2^+}{\sin\theta_2}
             \right] .
\end{align}
Here, the following notation has been introduced:
%
\begin{align}
\delta^3 = \delta(\omega_1 + \omega_2 - \varepsilon_- - \varepsilon_+)
            \times \\ \nonumber \times
            \delta(k_{1y} + k_{2y} - p^-_y - p^+_y) 
            \delta(k_{1z} + k_{2z} - p^-_z - p^+_z) ,
\end{align}
\begin{equation}
\label{qlll}
    q_j = \frac{\omega_j^2}{2m^2h}\sin^2\theta_j,  \qquad j=1,2,
\end{equation}
\begin{equation}
  T_j^\pm = e_{jx}\pm i e_{jy}, 
\end{equation}
$\vec e_j = (e_{jx}, e_{jy},e_{jz})$ are the polarization vectors of photons and $\phi_{1,2}$ are their angles of azimuth.

In resonance, we have $g_0^2 = E_n^2$, and the denominator of expression (\ref{G}) vanishes. 
To get rid of the divergence, we introduce width~$\Gamma$ of the intermediate state in accordance with the Breit–Wigner rule~\cite{Graziani95}:
%
\begin{equation} 
  E_n \rightarrow E_n - \frac{i}{2}\Gamma . 
\end{equation}

In calculating the total cross section, the integrals with respect to $d^2p^+$ and $dp^-_z$ can easily be evaluated using the properties of the $\delta$ function. 
However, expression (\ref{dsigma}) is independent of variable $p_{y}^{-}$; therefore, the result of integration with respect to $dp_{y}^{-}$ is factor $p_y$~\cite{Klepikov54}. 
In this case, the expression for the cross section acquires the term $p^-_yS/V$ in the form proposed in~\cite{Klepikov54}:
%
\begin{eqnarray}
  \frac{p^-_yS}{V} \rightarrow m^2h .
\end{eqnarray}

Finally, after simple computations, the resonant cross section of the process assumes the form
%
\begin{equation}
\label{smp}
  \sigma_{-+} = \sigma_0 (1 + \Xi_3)
                 \left[ 1 + u^2 + 2u\xi_2 - s^2\xi_3\right] ,  
\end{equation}
\begin{equation}
\label{smm}
  \sigma_{--} = \sigma_0 \frac{hl^+}{2}(1 - \Xi_3)
                 \left[ 1 + u^2 + 2u\xi_2 - s^2\xi_3\right] , 
\end{equation}
\begin{align}
\label{spp}
  \sigma_{++} = & \sigma_0 \frac{h}{2l^-} \left\{
                   (N^2\Xi_+ + n^2\Xi_-)(1 + u^2 + 2u\xi_2) + 
                   \right. 
                   \\ 
                 & + (N^2\Xi_+ - n^2\Xi_-)\xi_3s^2 +  
                   \\ \nonumber 
                 & \left.
                 + 2Nn\Xi_2 [2u + (1+u^2)\xi_2] - 
                   2Nn \Xi_1\xi_1 s^2 
                   \right\}, 
\end{align}
\begin{equation}
\label{spm}
  \sigma_{+-} = \sigma_0 \frac{h^2l^+}{4l^-}N^2(1 - \Xi_3)
                 \left[ 1 + u^2 + 2u\xi_2 + s^2\xi_3\right] ,
\end{equation}
where $\Xi_\pm = 1 \pm \Xi_3$, $u = \cos\theta_1$, $s = \sin\theta_1$, and
\begin{equation}
  \sigma_0 = \frac{\alpha^2\pi}{m^2}
              \sqrt{\frac{m}{\delta\omega}}
              \left( \frac{m}{\Gamma} \right)^2
              \frac{e^{-q_2}q_1^Nq_2^{l^++n}}{s^2(1-\cos\chi)}
              \frac{l^-!/l^+!}{N(N!n!)^2}.
\end{equation}
Here, $\Xi$ denotes the Stokes parameters of the hard photon, $\xi$ are these parameters for the soft photon, and the first and second subscripts in expressions (\ref{smp})--(\ref{spm}) indicate the sign of the electron and positron spin projections, respectively.

\begin{figure}
  \resizebox{\columnwidth}{!}{\includegraphics{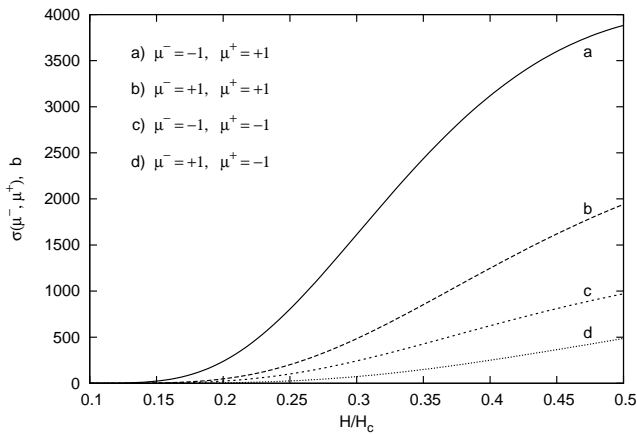}}
  \caption{Dependences of the cross section of production of a pair by nonpolarized photons on the magnetic field strength $h = H/H_c$. Landau levels are $l^- = 2$, $l^+ = 1$, and $\delta\omega = \omega_1 = mh$.}
  \label{S_h}
\end{figure}

Figure~\ref{S_h} shows the dependences of the resonant cross section of photoproduction on the magnetic field strength for various spin projections of the particles. 
It can be seen from expressions (\ref{smp})--(\ref{spm}) that the cross section has the highest order of magnitude for the production of a pair with spin directions $\mu^- = -1$ and $\mu^+ = +1$ as in the case of one-photon generation~\cite{Novak09}. 
This spin state corresponds to the minimal energy of interaction of magnetic moments of particles with the magnetic field. 
The reversal of the spin projection of each particle reduces the cross section by an order of magnitude in $h$.


The strong dependence of the cross section on the polarization of the hard photon is also worth noting.
In particular, for the normal linear polarization, we have $\Xi_3 = -1$, and cross section $\sigma_{-+}$ vanishes. 
It should be noted, however, that for correct comparison of cross section $\sigma_{-+}$ with other cross sections, it is necessary to calculate it with corrections in $h$ of the corresponding order of  magnitude. 
In the case of the linear polarization of photons, expressions (\ref{smp})--(\ref{spm}) are symmetric relative to the polar angle.

\end{document}